\documentclass[onecolumn]{article}

\usepackage{amsmath,amsfonts}
\usepackage{algorithmic}
\usepackage{array}
\usepackage{stfloats}
\usepackage{url}
\usepackage{graphicx}
\usepackage[caption=false,font=normalsize,labelfont=sf,textfont=sf]{subfig}
\usepackage[dvipsnames]{xcolor}
\usepackage[acronym]{glossaries}
\usepackage{multirow}
\usepackage{authblk}

\graphicspath{ {./} }

\newacronym{dwt}{DWT}{discrete wavelet transform}
\newacronym{dg}{DG}{distributed generation}
\newacronym{ann}{ANN}{artificial neural network}
\newacronym{emtp}{EMTP}{electromagnetic transient program}
\newacronym{ddbfl}{DDFL}{data-driven fault location}
\newacronym{twbfl}{TWBFL}{travelling wave based fault location}
\newacronym{ibfl}{IBFL}{impedance based fault location}
\newacronym{dps}{DPS}{distribution power system}
\newacronym{wt}{WT}{wavelet transform}
\newacronym{ml}{ML}{machine learning}
\newacronym{nn}{NN}{neural network}
\newacronym{lm}{LM}{Levenberg-Marquardt}
\newacronym{scg}{SCG}{scaled conjugate gradient}
\newacronym{pce}{PCE}{path correlation error}
\newacronym{phce}{PHCE}{phase correlation error}

\title{Data-Driven Ground-Fault Location Method in Distribution Power System With Distributed Generation}

\author[1]{Mauro Caporuscio}
\author[2]{Antoine Dupuis}
\author[1]{Welf L\"owe}
\affil[1]{Department of Computer Science and Media Technology, Linnaeus University}
\affil[2]{Department of Electrical Engineering, Uppsala University}
\date{}
\begin{document}

\maketitle

\begin{abstract}
The recent increase in renewable energy penetration at the distribution level introduces a multi-directional power flow that outdated traditional fault location techniques. To this extent, the development of new methods is needed to ensure fast and accurate fault localization and, hence, strengthen power system reliability. This paper proposes a data-driven ground fault location method for the power distribution system. An 11-bus 20 kV power system is modeled in  Matlab/Simulink to simulate ground faults. The faults are generated at different locations and under various system operational states. Time-domain faulted three-phase voltages at the system substation are then analyzed with discrete wavelet transform. Statistical quantities of the processed data are eventually used to train an Artificial Neural Network (ANN) to find a mapping between computed voltage features and faults. Specifically, three ANNs allow the prediction of faulted phase, faulted branch, and fault distance from the system substation separately. According to the results, the method shows good potential, with a total relative error of 0,4\% for fault distance prediction. The method is applied to datasets with unknown system states to test robustness.    
\end{abstract}

\section{Introduction}\label{sec:intro}
Power reliability is a very important property for any power grid. Reliability can be achieved through preemptive or fault location techniques. Preemptive techniques aim at avoiding outages by leveraging pathways and equipment redundancy, which requires large investments. Fault location techniques aim at decreasing the time for fault clearance. European networks provide a high level of reliability with a low number of faults and a short fault-clearance time (between 15 to 400 min per customer/year)~\cite{5719541}. However, since about 80\% of the faults observed in power networks occur at the distribution level, fault location in distribution networks has become an area of high interest for academic and industrial research, previously reserved for transmission networks~\cite{5719541}. 

Ground faults are very challenging to locate. Traditionally, the fault location process is initiated by customers, which notify an operator about an outage~\cite{FERREIRA2018696}. The trouble call only indicates one specific location compromised by the outage, and additional calls are needed to approximate the affected area. This information is then combined with knowledge about the network and positions of fault-clearing devices (e.g., Circuit Breaker) to identify possible fault locations. Relying on trouble calls as a fault indicator has several shortcomings: (1) customers tend to postpone the fault report, (2) reports are usually incomplete, (3) customers make mistakes and report false outages, and (4) night-time faults are less likely to be reported. Once a fault is verified and confirmed, a technician is usually dispatched to check, classify, and recover the fault~\cite{bahmanyar2017comparison}. This can be a tedious and resource-consuming task as the area in which the fault occurred might be very large, and the whole process usually depends on acquired knowledge about the area, such as previous experience and historical data~\cite{bahmanyar2017comparison}~\cite{FERREIRA2018696}. To this extent, data analysis can help to improve the accuracy of the fault location, reduce the time needed to fix the problem and optimize the resources~\cite{FERREIRA2018696}.

The main methods for fault location through data analysis can be categorized as~\cite{STEFANIDOUVOZIKI2022108031}: \textit{Impedance and Other Fundamental Frequency Component-Based Methods}  (IBFL) and \textit{High-Frequency Components and Traveling Wave Based Methods} (TWBFL).
IBFLs are widely used in distribution systems because of their cost-effectiveness.  The method computes multiple estimations by identifying a number of possible fault locations. In fact, the impedance is calculated starting from the measuring point and identifies all possible points in the network with an impedance equal to the assumed fault impedance. 
TWBFLs are based on the reflection and transmission of the fault-generated traveling waves on the faulted power network. Although the fault can be located with high accuracy, the implementation of such a technique is complex and more expensive than the implementation of impedance-based techniques.

Motivated by low cost and high potential, data-driven methods and associated machine learning techniques have been applied to many different fields and hence enjoy extensive literature and research nowadays. The field of fault location is not an exception to that. Recently, \acrfull{ddbfl} methods have been proposed to reduce the number of real-time calculations and, therefore, the computational load. In contrast to model-based techniques, which base fault localization on physical models, \acrshort{ddbfl} finds complex mappings based on observations, so-called training data. \acrshort{ddbfl} methods have been based on different machine learning (ML) algorithms, including \acrshort{ann} and Support Vector Machines~\cite{1413307,Yeetal2014}.
In this context, the main contributions of this paper are: 
\begin{enumerate}
    \item[(i)] Present a \acrshort{ddbfl} method adaptable to specific distribution networks.
    \item[(ii)] Compare the proposed method with traditional fault localization methods and other \acrshort{ddbfl} methods and highlight their difference and respective advantages.
\end{enumerate}

Our \acrshort{ddbfl} method (i) suggests the following steps, each detailed in a separate section. For a new distribution network, define a simulation model to generate (raw) training data as detailed in Section~\ref{sec:sim}. Instead of training the ML models with raw data, process them to extract useful information, which lets training converge faster. Section~\ref{sec:process} defines this data processing using \acrfull{dwt}. The transformed data is used for training ML models for the localization of faults anywhere in the specific distribution network. Section~\ref{sec:method} introduces our \acrshort{ann} models for \acrshort{ddbfl}. They are validated in Section~\ref{sec:validate}. 
Comparing our \acrshort{ddbfl} method with traditional and other \acrshort{ddbfl} methods (ii) is done in  the background and related work Section~\ref{sec:comp}; points in favor and against our method are further discussed in Section~\ref{sec:discuss}.
Finally,  Section~\ref{sec:conclude} concludes the paper and shows directions of future research.

\section{Background and Related Work}\label{sec:comp}
Fast fault recovery is provided by several types of fault location methods that have respective advantages, drawbacks, or limitations induced by their nature but also by their field of application. Concretely, a method suitable for one distribution power system (DPS) might be impractical for another one. This section reviews the methods suggested in the literature and spotlights their respective characteristics for application in distribution systems. Before, we briefly set the background of electrical power systems.

\subsection{Power Systems Background}

Power systems consist of three parts: generation, transmission, and distribution, where electrical power is first \emph{generated} at different power plants from which the bulk power is \emph{transmitted} across relatively long distances after passing through a step-up transformer. The electrical is then \emph{distributed} to commercial, industrial, and residential loads at different voltage levels. 

Carrying a tremendous quantity of electrical power over long distances, transmission lines' voltage levels must be high to avoid power loss. The power is then carried at a lower voltage through step-down transformers to be distributed to and consumed by customers. Distribution voltage levels can be divided into two groups: $4-35 kV$ for the primary distribution system and $120-240 V$ for secondary distribution systems at which the residential loads are connected. This occurs at primary distribution levels that \acrshort{dg} units can directly be connected to.

Differences in voltage level don’t impact fault location methods. However, the difference in topology between transmission and distribution systems does. While transmission systems are mainly purely radial, DPS consists of branches and tapped laterals dispersed over different types of areas. The electrical power is carried by the transmission system to the primary distribution substation through a step-down transformer. It is then carried to secondary distribution transformers via feeder conductors. DPS are also subject to changes in line composition and phase number, often leading to unbalanced operation. These characteristics make DPS more exposed to different fault sources such as wind, vegetation, lighting, storm, birds contact, and equipment failure. Also, it limits the use of methods that would be perfectly adapted to radial systems like transmission, as explained in section~\ref{sec:related}.

In addition to higher vulnerability and different topologies, modern DPS include dynamic load and DG from renewable sources. Adding a current source at this level of the overall system reshapes the power flow traditionally going from the transmission to load through distribution systems. As a result, power tends to flow dynamically within distribution systems, again limiting the use of classical methods for fault location.

\subsection{Ground Fault Location Methods}
\label{sec:related}

Traveling wave-based fault location (TWBFL) and impedance-based fault location (IBFL) fault location are commonly applied methods and give satisfying results. However, they both present non-negligible disadvantages, and higher performance must be reached to improve network stability further. To this end, emerging \acrfull{ddbfl} methods using \acrfull{ml} algorithms are investigated and promising results.

Some fault location algorithms require the determination of the faulted area first to be applied successfully. For instance, several methods give the distance between the distribution system substation and the fault as a result which raises problems of multiple possible locations when the network presents several branches, also called the multiple estimations problem. In this case, solutions for finding the correct faulted branch, i.e., the faulted area, must be designed and included in the algorithm. However, several independent means of fault area location exist, such as trouble call, fault indicator, or advanced metering infrastructure methods~\cite{bahmanyar2017comparison}.

\subsubsection{Traveling wave-based fault location}
When a fault occurs, high-frequency traveling waves of current and voltage are generated and propagate away from the fault point in both directions along the line toward its ends. The end of the line constitutes a circuit discontinuity in terms of wave propagation. Whenever a traveling wave reaches a transition point, it is divided into two secondary waves, one reflected and the other refracted. Their amplitudes or respective energy are attenuated due to medium property every time they meet a transition point and divide again until the subsequent refracted and reflected waves vanish, and the post-fault steady is reached~\cite{Aftabetal2020}. The foundation of TWBFL methods is based on exploiting these traveling waves: by analyzing their speed, the distance they travel, and the time they reach the relays located at the line terminals, one can determine the location where they originated.

As transmission systems are purely radial, their topology is particularly well suited for TWBFL methods. In fact, these methods were initially designed for transmission line fault location and mostly applied at this level, for which they are considered very mature and efficient techniques. On the contrary, DPSs have lateral and sub-branches. This specific topology involves shorter lines and more discontinuity, increasing the number and sampling rate of measuring devices to be able to distinguish superimposed traveling waves, making the implementation cost of this method higher at distribution level~\cite{bahmanyar2017comparison}. Nonetheless, TWBFL can still successfully be applied at distribution levels. The TWBFL adoption proposed in~\cite{Magnagoetal1998} was tested with simulation-generated traveling wave data. Two different tests gave errors relative to the total power line length of 0,45\% and 1,36\%, resp. The two-staged approach proposed in~\cite{Goudarzietal2015} determines the fault area first before accurate fault localization. For different fault locations, fault resistances, inception angles, and fault types, the smallest relative error is 0,34\%, and the largest is 3,31\% with an average of 1,69\%. Finally, an original and robust TWBFL method for single line to ground fault localization was proposed in~\cite{Chenetal2017}. Simulations involving different fault locations, fault resistances, inception angles, power line section area, and disturbance are run on a 34-bus system with and without two DG units to observe their impact on the method. Tests show that the method performs well even with DG units included. The mean relative error is 1,74\%.

TWBFL methods have both great advantages and drawbacks, which also depend on the applied method. Generally, the implementation and high cost of high sampling rate measurement devices, synchronization, and communication devices, as well as required knowledge about line configuration and parameters to compute propagation velocity, constitute non-negligible requirements and drawbacks. The limitation regarding the network topology is also a very important characteristic of this method. Systems that are not purely radial, just like DPS, present obstacles to the implementation of TWBFL methods. Also, discontinuity in power line material and size implies a change in wave velocity that increases errors. Moreover, it remains quite insensible to fault impedance. Finally, this method is mostly independent of network data and consequently insensitive to modeling errors.

\subsubsection{Impedance-based fault location}
Instead of transient signals, fundamental frequency root mean square (RMS) values of current and voltage during fault are exploited in IBFL methods. In most cases, phase voltages and currents are only available at the substation of the distribution system and constitute the required measurements to perform the various algorithms for fault location. The IBFL methods are based on circuit analysis theory and particularly the use of Kirchoff’s voltage and current law to estimate the fault distance. Starting from the first line section of the DPS, the faulted conditions of the next sections are found iteratively until the faulted section is reached, where the fault location can be estimated. This fault location highly depends on the impedance modeling of the overall system, such as self and mutual line impedance, load impedance as well as equivalent laterals. Hence, the accuracy of the method relies on how accurately the different system components are modeled and, therefore, highly relies on system knowledge. The main challenges associated with modeling are dynamic load profiles, distribution line heterogeneity, and distributed generation. Moreover, DG units have an intermittent and unpredictable generation that transforms the direction of power flow from radial to multi-directional and its magnitude from quasi-constant to variable within the distribution networks. Consequently, methods that claim to be suitable for high penetration of DG must consider these changes. Hence, one IBFL method very optimized for one type of DPS might not be adapted for a different one~\cite{MoraFlorezetal2008}.

In~\cite{NB2010}, the authors present a method to calculate ground fault location adapted to unbalanced systems with DG. The problem is solved using RMS values of voltage and current during the fault available at the system substation. The adaptation of the method to the presence of DG shows good results with an average relative error of 0,2\%. An adaption of the IBFL method in~\cite{NB2010} for unbalanced DPS considering shunt admittance and dynamic load is explained in~\cite{Ferreiraetal2012}. This method is tested on a 13-buses system. The results show the effect of the load profile matching algorithm, which allows a relative error of only 0,92\% for a load variation of 50\% compared to the initial load profile. When the load doesn't vary, this method gives a relative error of 0,12\%. Another related method was proposed on a 6-buses DPS~\cite{Menchafouetal2015}. In contrast to the methods discussed earlier, the authors use RMS voltage and current measurements at every bus, which simplifies the method formulation. Similar to~\cite{NB2010}, the fault current is approximated to be able to calculate the fault location. A DG unit was also added to test the impact of DG on the method. Results are promising, with a maximal relative error of 0,124\% showing the robustness of the method. The average relative error is as low as 0,024\% due to the larger availability of real measurements within the system. 

Overall, IBFL methods are simple to implement and relatively low cost, notably without the need for high-frequency measurement devices. They constitute the reference for fault location at distribution levels. In that sense, significant progress has been made to overcome the main challenges brought by the required accurate modeling of DPSs. Indeed, solutions and algorithms to model dynamic load, laterals as well as penetration of distributed generation have been designed. As an important drawback, system knowledge is mandatory for IBFL methods to be applied since they are based on the equivalent impedance of the system, such as line and load, and modeling of other system components. Another requirement is the measurement of faulted RMS voltages and currents at the substation.

\subsubsection{Data-driven methods}
are based on the use of a large amount of data containing knowledge about the system under fault occurrence. The field of \acrshort{ml} underwent major developments during the past recent years, being applied to many different areas, including the field of fault location. Techniques attempt to learn or train general fault location functions (\acrshort{ml} models) from examples of fault locations and corresponding observations. \acrshort{ml} models can predict either continuous or discrete values, corresponding to classification and regression problems, resp. The arguments are called features, and their corresponding outputs are either called targets for regression and categories for classification problems. A part of the dataset is used for training, and the remaining part to testing the trained model. Since the model is fit the training dataset, the testing dataset enables an unbiased evaluation of the trained model. While \acrshort{ml} techniques can find highly complex nonlinear functions, their interpretation can be challenging, if not impossible, for human logic. Consequently, finding the best \acrshort{ml} model is often an error and trial process.

The selection of the dataset plays a key role in guaranteeing that the trained model generalizes well to a new, unknown input. That is, the chosen dataset and its features should be similar to the potential future input vector that this model will be used on. When a fault occurs, the power system can operate under various conditions, and the fault has different parameters. Operating conditions mainly include the rate of DG penetration and load values; fault parameters are the fault impedance, the faulted phases, and the angle of the voltage phase when the fault occurs, called the inception angle. The unique combination of different system operating conditions and fault parameters for a given fault occurrence is called a fault scenario. To be generalizable, the dataset should be associated with realistic and various fault scenarios. In the case of the power system fault location problem, transient or steady-state three-phase voltage and/or current measured at one or more locations within the system during the fault are most often taken as the features for training an \acrshort{ml} model. Naturally, the location of the fault is the target of the model. Measurements of voltage and current should be done under various fault scenarios and at different locations within the system, enabling the \acrshort{ml} model to learn the mapping of these signals to the location of the fault.

The accuracy of \acrshort{ddbfl} methods highly depends on the degree of detail and exactitude of the power system modeling. To perform well, the system model used to generate the dataset should be as close as possible to the real system on which the method will be applied in real life. In practice, when a real fault occurs, measurements are done and provide the input to the \acrshort{ml} model trained on the training data generated by the system model. As a result, the \acrshort{ml} model predicts the fault location.

In~\cite{Yeetal2014}, a method to locate ground fault uses support vector regression to compute the fault location. Traveling wave data recorded at the substation are used as raw data. It is then processed with modal transformation on which discrete wavelet transform of scale from 1 to 6 is applied. The arrival time delay between modal components and peak amplitude ratio of modal components wavelet coefficients are used as input features to train the \acrshort{ml} model. This data is generated for a balanced long distribution system that includes two laterals. Faults are simulated every 0,5 km. The number of simulations performed is as low as 40 since the correlation between chosen features and fault location is easily established and almost proportional. In terms of results, this method predicts fault location with a mean relative error of 6,2\% after testing the method with 17 samples. 

The method proposed in~\cite{GP2016} uses different variants of \acrshort{ann}s to classify and locate faults within a transmission system. The data is recorded at the faulted transmission line terminal and contains measurements for the half-cycle duration of three-phase current, voltage, and power phasors. Samples of each signal are arranged in individual matrices, and their respective determinant is computed, serving as input features for the \acrshort{ann}s. In total, 3'000 fault scenarios are simulated at 27 different fault locations. Fault type classification \acrshort{ann} uses 32 input features, whereas only 6 are used for fault location \acrshort{ann}. Networks are trained with a training dataset of 400 simulations, and the remaining 80'600 simulations are used for testing. Even with this small training dataset, faults are classified with an accuracy of 99,8\% and location is predicted with a mean relative error of 6,3\%.

The method described~\cite{PS2010} exploits the relation between the energy spectrum of voltage transient traveling waves recorded at the substation and fault location. The energy spectrum is found by using the discrete wavelet transform at scale 8 to 4'096, and input features are the energy content of each scale, corresponding to 10 input features. To predict the fault location, an \acrshort{ann} is chosen. The training is done with 3'474 samples of different fault scenarios in a distribution system that contains 5 laterals. Applied to the test data, the method has a mean relative error of 0,5\%.

\acrshort{ddbfl} methods are flexible since they can be applied to any power system type, architecture, and topology as long as this system can be modeled accurately enough using \acrshort{emtp}s. The accuracy of the built model directly impacts the precision of the method: the whole system and its numerical model should behave similarly in faulted simulation. Hence, limitations might be found when building a very detailed model, leading to a computation burden when the faults are simulated for data generation. Moreover, in real life, various disturbances of different natures that cannot be anticipated or modeled in the numerical version could potentially lead to more significant errors in fault location prediction. Since \acrshort{ddbfl} methods rely on accurate power system modeling, for any change in the network topology or else, the numerical model should be updated, data generated once again, and the ML model retrained.

\section{Dataset Generation}\label{sec:sim}
The major obstacle to employing \acrshort{ml} in practice is the lack of existing datasets. Indeed, available data on faulted power systems are neither large enough nor pertinent enough. To solve this problem, \acrfull{emtp} systems allow the modeling of power systems, the simulation of ground faults as well as the measurement of voltage or current signals everywhere in the system. Consequently, a large and relevant dataset of faulted voltage and current can be generated with the help of simulation, enabling the development of \acrshort{ddbfl} methods.

When a fault occurs, high-frequency traveling waves of current and voltage are generated and propagate away along the line toward its ends. Therefore, voltage and current waveforms undergo a discontinuity where high-frequency transients are added to steady-state sinusoids until a new post-fault steady-state is reached. These transients, their frequency, and their magnitude are dependent on the system states and the fault locations. To this extent, the transient three-phase voltages at the system substation are chosen to constitute the raw dataset of the data-driven method.

\begin{figure}[t]
\begin{center}
\includegraphics[width = .8\columnwidth]{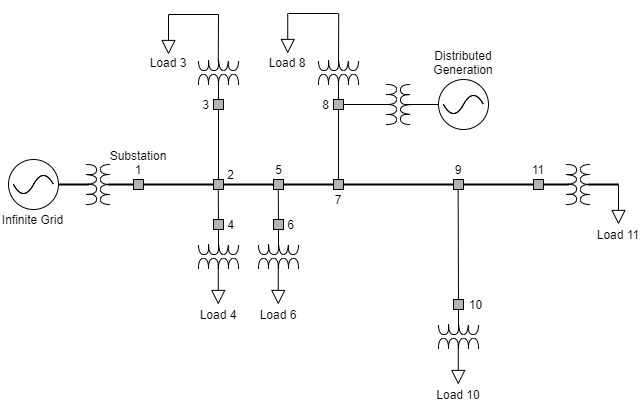}
\end{center}
\caption[Distribution power system test model structure.]{Distribution power system test model structure.}
\label{DPSstructure}
\end{figure}

The modeling of \acrfull{dps} enables the on-demand generation of large datasets required for the development of \acrshort{ddbfl} methods. To generate three-phase faulted voltages at different locations and under different system states, a \acrshort{dps} model shown in Figure~\ref{DPSstructure} is developed in Simulink$^\mathtt{TM}$. It is a balanced 11-bus 20 kV \acrshort{dps} to which a distributed generation unit modeled by a synchronous generator is added. Its topology and network data are inspired from~\cite{PS2010} and given in appendix \ref{ap:DPSdata}.

To be relevant, the dataset should map well the possible system states, that is to have fault occurrence under various system conditions. Hence, a total of 168 different fault scenarios are chosen for fault occurrence. Two percentage of \acrshort{dg} penetration in the system (DG\%: 10\% and 50\, seven faulted phase(s) ($F_{ph}$: a, b, c, ab, ac, bc, abc), four fault impedance ($Z_f$: 0,01, 0,1, 1 and 10$\Omega$), and three inception angles ($\theta_i$: 45°, 90° and 135°) are chosen. To reduce computation time, loads remain unchanged in all scenarios. In addition, faults are simulated every 500 meters, giving a total of 38 fault locations, and bringing the total number of fault simulations to 6384. Fault occurrence time is taken at $0,025$ s and the sampling frequency is $0,67$ MHz; 

\begin{figure*}[t]
         \centering
         \subfloat[$DG = 10\%$ ; $F_{ph}=a$ ; $Z_f=0,01\Omega$ ; $\theta_i=45$ ; $x=500$m on path 1]{%
              \includegraphics[width=0.48\textwidth]{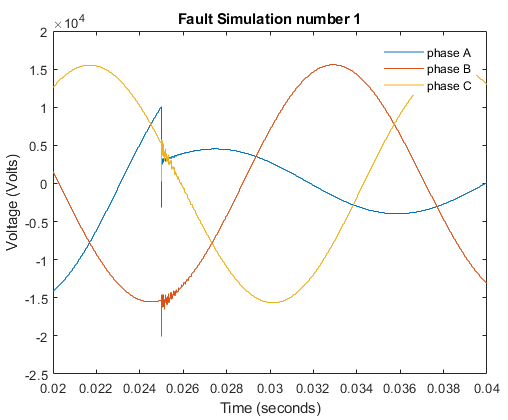}
              \label{fig:rawdata1}
              }
         \hfill
         \subfloat[$DG = 50\%$ ; $F_{ph}=ab$ ; $Z_f=0,1\Omega$ ; $\theta_i=90$ ; $x=3000$m on path 1]{%
              \includegraphics[width=0.48\textwidth]{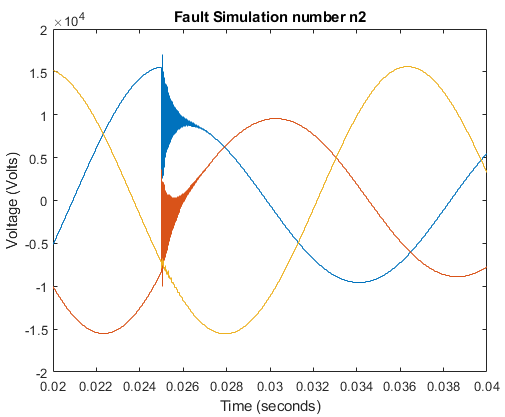}
              \label{fig:rawdata3}
              }
        \\
         \subfloat[$x=2000$m on path 1]{%
              \includegraphics[width=0.48\textwidth]{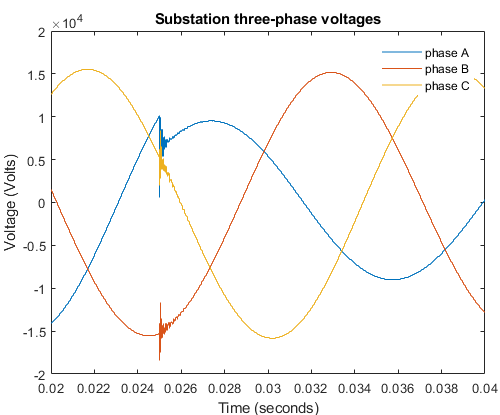}
              \label{fig:locimpact2}
              }
         \quad
         \subfloat[$x=8000$m on path 1\label{fig:locimpact8}]{%
              \includegraphics[width=0.48\textwidth]{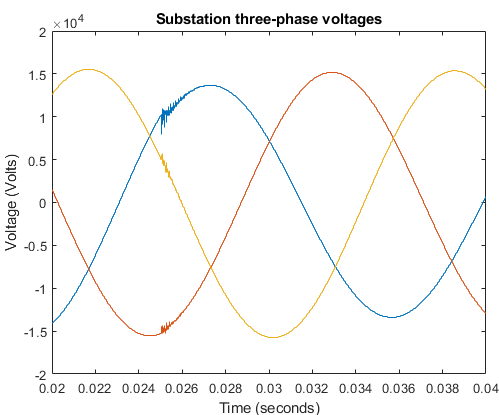}}
              
\caption{Three-phase voltages recorded at the system substation under various fault scenarios and different fault locations.}
\label{rawdata}
\end{figure*}

Figure~\ref{rawdata} shows four simulations of faulted three-phase voltages under different fault scenarios. In particular, two-phase a-to-ground faults occurring under the same system states but located at (c) 2000 m and (d) 8000m on path 1 put into light the impact of the fault location on the three-phase faulted voltages.

In addition, the three time-domain modal voltages $V_0$, $V_1$ and $V_2$ are computed using modal transform. They form, together with the three-phase voltages, the raw dataset of the method.

\section{Data Processing}\label{sec:process}
After simulation, the generated raw dataset must be processed to extract useful information. The use of \acrshort{dwt} is described in this section.

\subsection{Discrete Wavelet Transform} \label{sec:sig pro}

Transient signals are located in time and so cannot be captured by classical signal processing methods, such as the Fourier transform, which is used for stationary signals, or short-time Fourier transform, which has fixed time and frequency resolutions. These shortcomings are solved by \acrfull{wt} which allows simultaneous time-frequency analysis of signals at different time-frequency resolutions. 

The \acrshort{wt} is based on a mother-wavelet function $\psi(t)$ that can be stretched or compressed in time with the use of scaling parameter $a$. Each unique scaled wavelet $\psi_{a,b}(t)$, called daughter-wavelets, is shifted in time along the signal $s(t)$ via the shifting parameters $b$ the signal. The mother-wavelet must be finite and well localized in time and satisfy specific conditions, described in \cite{4a}. Daubechies-4 (db4) is chosen for this study. The continuous wavelet transform produces coefficients given as follows:
\begin{equation}
\begin{split}
& C_{a,b}=\int_{-\infty}^{+\infty}s(t)\psi^*_{a,b}(t)dt\\
& a>0,-\infty<b<+\infty.
\label{CWT}
\end{split}
\end{equation}
Where $\psi^*_{a,b}(t)$ is the conjugate of the daughter wavelet defined by (\ref{daughter})
\begin{equation}
\psi_{a,b}(t)=\frac{1}{\sqrt{a}}\psi\left(\frac{t-b}{a}\right).
\label{daughter}
\end{equation}
The wavelets coefficients are obtained for a center pseudo-frequency of the wavelet decomposition that can be expressed as: 
\begin{equation}
F=\frac{f_c f_s}{s}
\label{centerfreq}
\end{equation}
with $f_c$ the center frequency of the daughter wavelet and $f_s$ the sampling frequency of the signal. Because the scaling parameter is inversely proportional to the wavelet frequency $f_c$, the use of different scales $a$ allows the extraction of different frequencies from the original signal. This way, the signal can be analyzed in different frequency bands. The energy content of each scale $a$ or frequency, $E_{wave}$ is defined by (\ref{energy}) below:
\begin{equation}
E_{wave}(a)=\sum_{n=0}^{N-1}(C(a,nT_s))^2.
\label{energy}
\end{equation}
The numerical implementation of the continuous \acrshort{wt} on a sampling signal $s(t)$ is as follows (according to \cite{4a}):
\begin{gather}
C_{a,iT_s}=\frac{T_s}{\sqrt{a}}\sum_{n=0}^{N-1}\psi^*\left(\frac{(n-i)T_s}{a}\right)s(nT_s)\label{coef}\\
i=0,1,2,...,N\nonumber
\end{gather}
where $N$ is the number of signal samples and $T_s$ is the sampling period.

\begin{figure}[h]
\begin{center}
\includegraphics[width=.8\columnwidth]{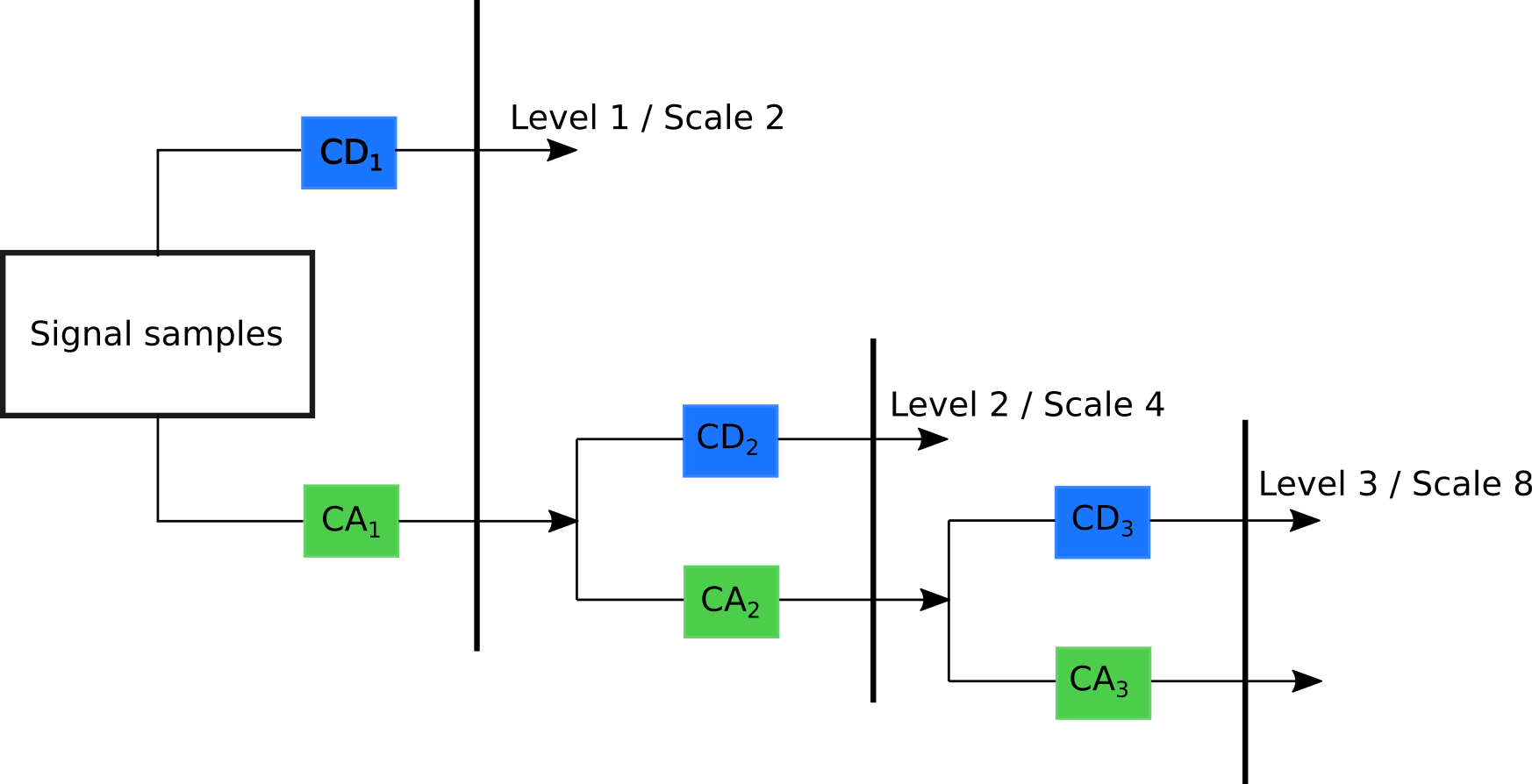}
\end{center}
\caption[Discrete wavelet transform with multi-level decomposition]{Discrete wavelet transform with multi-level decomposition where $j=1,2,3$.}
\label{discretewt}
\end{figure}

From (\ref{coef}), one can deduce that the number of wavelet coefficients is equal to the number of samples times the number of scales which can lead to computational burdens. The \acrshort{dwt} allows quicker signal processing by using dyadic scaling and shifting. This implies discrete scaling and shifting parameters shown in (\ref{dyadic}). 
\begin{equation}
\begin{split}
& a=2^j\\
& b=2^jm
\label{dyadic}
\end{split}
\end{equation}
where $j$ and $m$ are integer numbers. Hence, the discrete daughter wavelets are as follows
\begin{equation}
\psi_{j,m}(t)=\frac{1}{\sqrt{2^j}}\psi\left(\frac{t-m2^j}{2^j}\right).
\label{discretedaughter}
\end{equation}
The implementation of \acrshort{dwt} decomposes the signal using high-pass and low-pass filters, giving high-pass and low-pass sub-banks. The high-pass and low-pass sub-banks are populated by detail coefficients $CD$ and approximation coefficients $CA$, respectively. If $j>1$, a multi-level decomposition occurs as described in Figure~\ref{discretewt}. After each level of decomposition $j$ with associated scale $2^j$, the details coefficients become inputs of the next level filtering with scale $2^{j+1}$ giving high-pass and low-pass sub-banks of level $j+1$.
The covered frequency bands for each level decomposition $j$ are expressed below (according to \cite{Goudarzietal2015}):
\begin{equation}
\begin{split}
& CD_j:[2^{-(j+1)}f_s,2^{-j}f_s]\\
& CA_j:[0,2^{-(j+1)}f_s]
\end{split}
\label{freqrge}
\end{equation}
Corresponding values for $f_s=0.67$ MHz are presented in Table~\ref{DWTfreq} for $j=[1;8]$.
\begin{table}
\footnotesize{
\begin{center}
\begin{tabular}{|c|c|c|c|c|}
\hline
{j} & {CD} & {Frequency band} & {CA} & {Frequency band}\\
\hline
{$1$} & {$CD_1$} & {$167.5-335$ kHz} & {$CA_1$} & {$0-167.5$ MHz}\\
{$2$} & {$CD_2$} & {$83.7-167.5$ kHz} & {$CA_2$} & {$0-83.7$ MHz}\\
{$3$} & {$CD_3$} & {$41.9-83.7$ kHz} & {$CA_3$} & {$0-41.9$ kHz}\\
{$4$} & {$CD_4$} & {$20.9-41.9$ kHz} & {$CA_4$} & {$0-20.9$ kHz}\\
{$5$} & {$CD_5$} & {$10.5-20.9$ kHz} & {$CA_5$} & {$0-10.5$ kHz}\\
{$6$} & {$CD_6$} & {$5.2-10.5$ kHz} & {$CA_6$} & {$0-5.5$ kHz}\\
{$7$} & {$CD_7$} & {$2.6-5.2$ kHz} & {$CA_7$} & {$0-2.6$ kHz}\\
{$8$} & {$CD_8$} & {$1.3-2.6$ kHz} & {$CA_8$} & {$0-1.3$ kHz}\\
\hline
\end{tabular}
\caption{Frequency band values of multi-level DWT analysis for levels $j=[1;8]$}
\end{center}
}
\label{DWTfreq}
\end{table}

The higher the decomposition level is, the fewer the samples and hence, wavelet coefficients. Therefore, high-frequency resolution but low time resolution is obtained, and the other way around for small-level decomposition. 

\begin{figure*}[!t]
\begin{center}
\includegraphics[width=\textwidth]{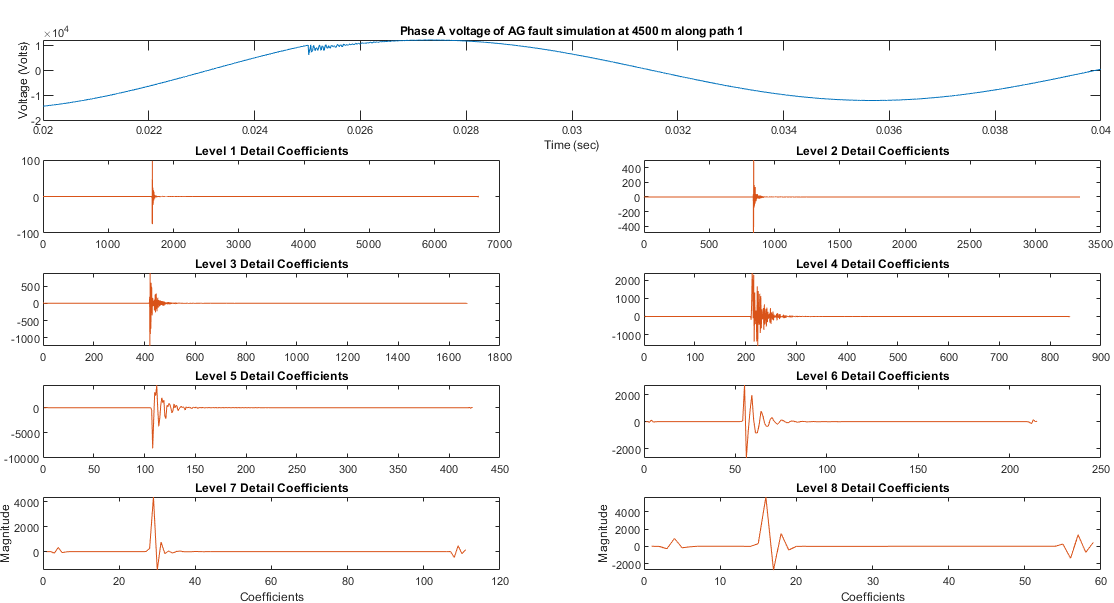}
\end{center}
\caption[Detail wavelet coefficients of phase A voltage.]{Detail wavelet coefficients of phase A voltage for a phase A-to-ground fault at 4500 m along path 1.}
\label{CDrawdata}
\end{figure*}

The dataset is processed with \acrshort{dwt} until level 8. The computation is carried out on MATLAB$^{TM}$ with the help of the Wavelet toolbox. The decomposition of a three-phase voltages sample is shown in Figure~\ref{CDrawdata}. 

\subsection{Relevant Features Extraction}\label{featextract}

After having processed voltage phases and modes signals, relevant features should be extracted to constitute the final dataset. In this paper, statistical quantities of the processed dataset are used as input to the neural networks. These are standard deviation $(std)$, variance $(var)$, third central moment $(mom3)$, skewness $(skn)$, and mode $(mode)$ and are applied to the following data for every simulation:
\begin{itemize}
    \item Three-phase time-domain voltages: $V_a$, $V_b$ and $V_c$
    \item Time-domain modal voltages: $V_0$, $V_1$ and $V_2$
    \item Level 8 detail and approximation wavelet coefficients of each phase voltages: $V_aCD_8$, $V_bCD_8$, $V_cCD_8$, $V_aCA_8$, $V_bCA_8$ and $V_cCA_8$
    \item Level 8 detail and approximation wavelet coefficients of each modal voltages: $V_0CD_8$, $V_1CD_8$, $V_2CD_8$, $V_0CA_8$, $V_1CA_8$ and $V_2CA_8$
\end{itemize}
In addition, frequency energy contents, $E_{wave}(V_aCD_8)$,$...$, $E_{wave}(V_2CA_8)$, are also taken as input features.

\begin{figure}[t]
\begin{center}
 \includegraphics[width=0.6\columnwidth]{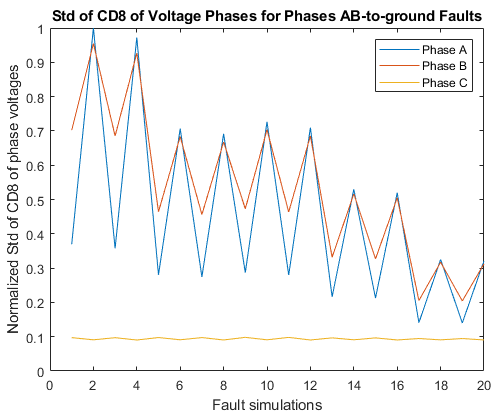}    
\end{center}
         \caption{Correlation between faulted phase and standard deviation.}
         \label{fig:relphase}
\end{figure}

Relation between these inputs and the networks' targets can be directly observed. Indeed, correlation with faulted phase(s) classification is shown in Figure~\ref{fig:relphase} where the standard deviation of $CD_8$ of each phase is plotted for 20 different simulations having the same faulted phases. One can observe that there is a discernible difference between healthy (c) and faulty phases (ab).  

\begin{figure}[h]
\begin{center}
\includegraphics[width=.6\columnwidth]{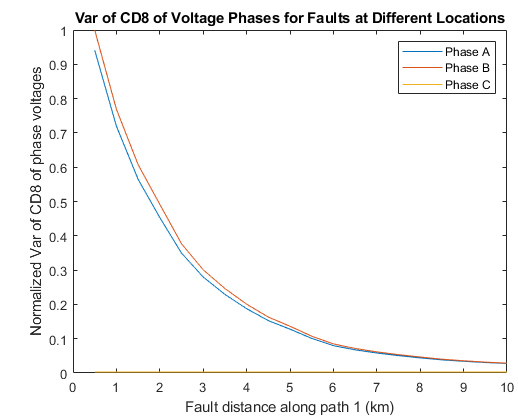}    
\end{center}
         \caption{Correlation between fault distance and variance.}
         \label{fig:reldistance}
\end{figure}

Similarly, Figure~\ref{fig:reldistance} shows the effect of fault distance by giving the variance of $CD_8$ for ab-to-ground fault simulation that differs only by the fault distance along path 1. It is visible that the variance of $CD_8$ of the faulted phases decreases with distance.

\begin{figure}[h]
\centering
         \includegraphics[width=.6\columnwidth]{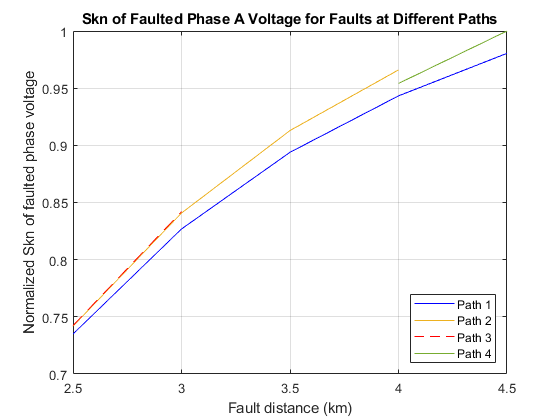}
         \caption[Correlation between faulted path and skewness.]{Correlation between faulted path and skewness.}
         \label{fig:relpath}
\end{figure}

The faulted path is, however, harder to identify from the available features. This is attested in Figure~\ref{fig:relpath}. It indicates the skewness of faulted phase voltages from simulations with the same fault scenario but different locations in distance and path. One can notice that faults occurring at the same distance but on different paths aren't well dissociated for paths 2 and 3. Nevertheless, paths 1, 2, and 4 can be distinguished quite well.

Only the statistical quantities having a noticeable correlation with the networks' targets are kept as input features for the fault location method. They are referred to later in the next section.

\section{Fault Location Models}\label{sec:method}
After the determination of the relevant features, they are fed as input to three different sets of \acrshort{ann}s. Each set has three distinct prediction targets, namely the faulted phase(s), the fault distance, and the faulted path. The fault location methodology using these different sets is described in this section.

\subsection{Artificial Neural Networks}

\begin{figure}[h]
\begin{center}
\includegraphics[width=\columnwidth]{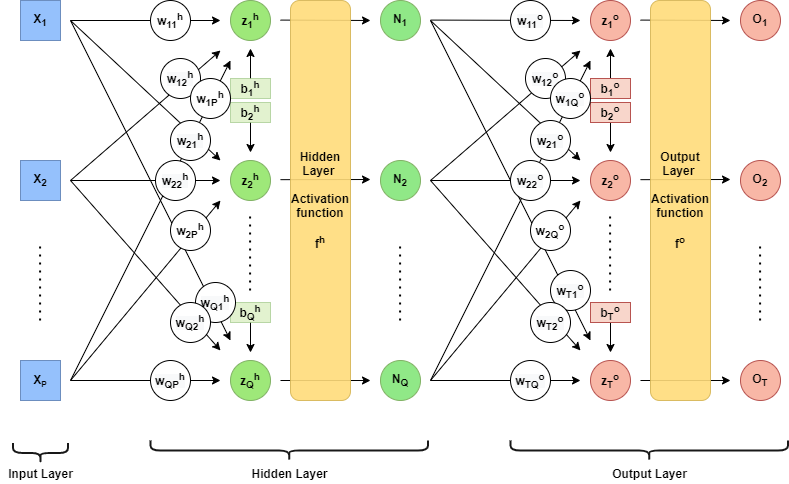}
\end{center}
\caption{Artificial neural network architecture.}
\label{fig:ANNstruct}
\end{figure}

\acrshort{ann}s are used for their strong predictive power as well as their high versatility. Given enough training data, an \acrshort{ann} can find \emph{any} mapping between input and output where other conventional methods can't. The structure of an \acrshort{ann} is shown in Figure~\ref{fig:ANNstruct}.

Feed-forward neural networks are chosen to predict the fault distance, whereas for faulted phase and path classification, pattern recognition type networks are utilized. The \acrfull{lm} and the \acrfull{scg} optimization algorithms are chosen to train the feed-forward and pattern recognition neural network, respectively. Every individual \acrshort{ann}s deployed in the method is described in Table~\ref{anns}, where $P$ is the number of input features, $Q$ is the number of hidden neurons, and $T$ is the number of outputs. $f^h$ and $f^o$ are the hidden and output layer activation function, respectively. Each \acrshort{ann} is trained with the different input features determined in the previous Section~\ref{sec:process} according to their relevancy for the respective \acrshort{ann}'s prediction. These inputs are referred to in Appendix \ref{ap:DPSdata}. Optimum weights and biases values aren't presented here for clarity purposes (but can be made available on demand). Finally, the structure, that is, the number of hidden neurons of each individual \acrshort{ann}s, was based on a trial-and-error approach for which $10 \leq Q \leq 50$, and hence differs between \acrshort{ann}s. The available datasets are divided as follows: 70\% is used for training, and the remaining 30\% is later used for testing and validation.

\begin{table}
\scriptsize{
\begin{center}
\begin{tabular}{|c|c|c|c|c|c|c|}
\hline
{\textbf{ANN}} & {\textbf{Data size}} & {\textbf{P}} & {\textbf{Q}} & {$f^h$} & {\textbf{T}} & {$f^o$}\\
\hline
\hline
\multicolumn{7}{|c|}{\textit{Faulted Phase Classification}}\\
\hline
{Ph} & 6384 & {15} & {10} & {tansig} & {7} & {softmax}\\
\hline
\hline
\multicolumn{7}{|c|}{\textit{Fault Distance Regression}}\\
\hline
{D-a} & 912 & {23} & {10} & {tansig} & {1} & {$f(x)=x$}\\
{D-b} & 912 & {23} & {20} & {tansig} & {1} & {$f(x)=x$}\\
{D-c} & 912 & {23} & {30} & {tansig} & {1} & {$f(x)=x$}\\
{D-ab} & 912 & {38} & {10} & {tansig} & {1} & {$f(x)=x$}\\
{D-ac} & 912 & {38} & {30} & {tansig} & {1} & {$f(x)=x$}\\
{D-bc} & 912 & {38} & {10} & {tansig} & {1} & {$f(x)=x$}\\
{D-abc} & 912 & {45} & {30} & {tansig} & {1} & {$f(x)=x$}\\
\hline
\hline
\multicolumn{7}{|c|}{\textit{Faulted Path Classification}}\\
\hline
{Pa-a$h_1$} & 408 & {30} & {50} & {tansig} & {4} & {softmax}\\
{Pa-b$h_1$} & 408 & {30} & {50} & {tansig} & {4} & {softmax}\\
{Pa-c$h_1$} & 408 & {30} & {30} & {tansig} & {4} & {softmax}\\
{Pa-ab$h_1$} & 408 & {37} & {50} & {tansig} & {4} & {softmax}\\
{Pa-ac$h_1$} & 408 & {37} & {50} & {tansig} & {4} & {softmax}\\
{Pa-bc$h_1$} & 408 & {37} & {30} & {tansig} & {4} & {softmax}\\
{Pa-abc$h_1$} & 408 & {26} & {30} & {tansig} & {4} & {softmax}\\
{Pa-a$h_2$} & 504 & {30} & {50} & {tansig} & {3} & {softmax}\\
{Pa-b$h_2$} & 504 & {30} & {50} & {tansig} & {3} & {softmax}\\
{Pa-c$h_2$} & 504 & {30} & {30} & {tansig} & {3} & {softmax}\\
{Pa-ab$h_2$} & 504 & {37} & {50} & {tansig} & {3} & {softmax}\\
{Pa-ac$h_2$} & 504 & {37} & {50} & {tansig} & {3} & {softmax}\\
{Pa-bc$h_2$} & 504 & {37} & {30} & {tansig} & {3} & {softmax}\\
{Pa-abc$h_2$} & 504 & {26} & {30} & {tansig} & {3} & {softmax}\\
\hline
\end{tabular}
\caption{Artificial neural networks structure}
\end{center}
}
\label{anns}
\end{table}

\subsection{Faulted Phase Prediction}

The faulted phase is computed using a single \acrshort{ann}, namely Ph, that hence predicts the fault type: $ag$, $bg$, $cg$, $abg$, $acg$, $bcg$ or $abcg$. It has 15 input features and uses the 6384 samples generated dataset for training and testing. 

\subsection{Fault Distance Prediction}

The fault distance prediction strategy takes advantage of the faulted phase classification results: depending on the fault type, one of the seven fault distance \acrshort{ann}s returns the fault distance $x$ from the substation as output. If the faulted phase is found to be phase a, D-a is used, and similarly for other faulted phases. Each of them is trained exclusively with input data related to the faulted phase(s) they are specialized on. Consequently, only a ratio of the generated dataset can be used for their training, that is, 912 samples out of 6384 for each of the seven \acrshort{ann}s.

\subsection{Faulted Path Prediction}

\begin{figure}[!t]
\centering
\includegraphics[width=.8\columnwidth]{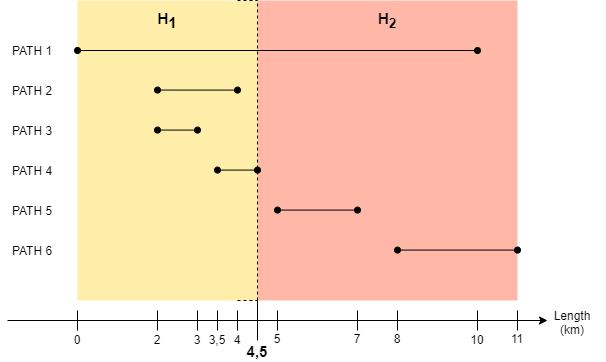}
\caption{Faulted path classification ANNs groups $h_1$ and $h_2$.}
\label{fig:H1H2}
\end{figure}

To avoid multiple location estimation problems, the faulted path must be determined too. To this extent, the method uses 14 different trained \acrshort{ann}s: in addition to having one \acrshort{ann} per fault type, they are further divided into two groups $h_1$ and $h_2$. The groups $h_1$ for fault distance occurring at a distance of $x\leq 4500$ meters and $h_2$ for $x>4500$ meters. This is illustrated in Figure~\ref{fig:H1H2}. This allows us to reduce the number of categories that can be predicted. The $h_1$ labeled \acrshort{ann}s can predict four different paths, from 1 to 4, and only three for the others, paths 1, 5, and 6. Thus, the sample set size related to each \acrshort{ann} is even smaller and goes down to 408 and 504 samples for $h_1$ and $h_2$ groups, respectively. 

\begin{figure}[!t]
\centering
\includegraphics[width=.5\columnwidth]{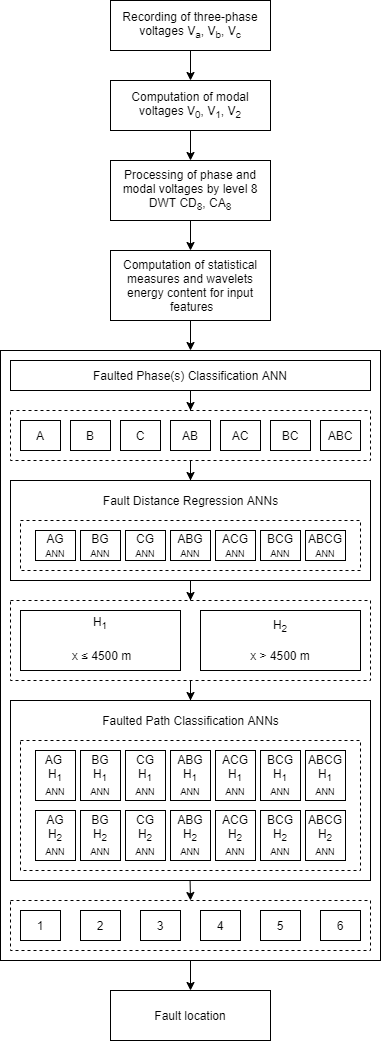}
\caption{Fault location algorithm using optimized workflow.}
\label{fig:flalgo}
\end{figure}

The algorithm of the method, from faulted phase to faulted path classification, including data processing and feature extraction steps, is summarized in Figure~\ref{fig:flalgo}.

\section{Method Validation}\label{sec:validate}
After training the different networks involved in the method algorithm, a validation test is carried out of the remaining 30\% of the different samples set. This section presents the theoretical results of the method, summarized in Table~\ref{accANN}.

\begin{table}[t]
\footnotesize{
\begin{center}
\begin{tabular}{|c|c|c||c|}
\hline
{\textbf{ANN}} & {\textbf{Classification}} & {\textbf{Mean} $e_{rel}$} & {\textbf{Total}}\\
\hline
\hline
\multicolumn{4}{|c|}{\textit{Faulted Phase Classification}}\\
\hline
{Tfph} & {100\%} & {n/a} & {100\%}\\
\hline
\hline
\multicolumn{4}{|c|}{\textit{Fault Distance Regression}}\\
\hline
{Ofd-A} & {n/a} & {0,25\%} & \multirow{7}{2.5em}{0,40\%}\\
{Ofd-B} & {n/a} & {0,58\%} &\\
{Ofd-C} & {n/a} & {0,43\%} &\\
{Ofd-AB} & {n/a} & {0,3\%} &\\
{Ofd-AC} & {n/a} & {0,55\%} &\\
{Ofd-BC} & {n/a} & {0,32\%} &\\
{Ofd-ABC} & {n/a} & {0,40\%} &\\
\hline
\hline
\multicolumn{4}{|c|}{\textit{Faulted Path Classification}}\\
\hline
{Ofp-A$H_1$} & {73,77\%} & {n/a} & \multirow{14}{3em}{75,15\%}\\
{Ofp-B$H_1$} & {68,85\%} & {n/a} &\\
{Ofp-C$H_1$} & {67,21\%} & {n/a} &\\
{Ofp-AB$H_1$} & {77,6\%} & {n/a} &\\
{Ofp-AC$H_1$} & {65,5\%} & {n/a} &\\
{Ofp-BC$H_1$} & {65,5\%} & {n/a} &\\
{Ofp-ABC$H_1$} & {62,29\%} & {n/a} &\\
{Ofp-A$H_2$} & {98,68\%} & {n/a} &\\
{Ofp-B$H_2$} & {85,52\%} & {n/a} &\\
{Ofp-C$H_2$} & {76,31\%} & {n/a} &\\
{Ofp-AB$H_2$} & {62,29\%} & {n/a} &\\
{Ofp-AC$H_2$} & {80,26\%} & {n/a} &\\
{Ofp-BC$H_2$} & {93,42\%} & {n/a} &\\
{Ofp-ABC$H_2$} & {75\%} & {n/a} &\\
\hline
\end{tabular}
\caption{ANNs prediction results}
\end{center}
}
\label{accANN}
\end{table}

\subsection{Fault Location Results}

\subsubsection{Faulted Phase Classification}

After testing, the faulted phase can be predicted with an accuracy of 100\%, showing the great potential of the method for faulted phase classification. 

\subsubsection{Fault Distance Regression}

The performance of the method for fault distance prediction is evaluated with the mean relative error computed with equation (\ref{relativeer}). 
\begin{equation}
e_{rel}=\frac{|D_{cal}-D_{real}|}{L}
\label{relativeer}
\end{equation}
where $D_{cal}$ and $D_{real}$ are the calculated and real fault distance from the substation respectively. For the distribution system used, the maximum distance from the system substation is $L=11$ km. 

As a result, the combined mean relative errors of the seven fault distance prediction \acrshort{ann}s is as low as 0,40\%, showing here again considerable potential.

\subsubsection{Faulted Path Classification}

In terms of faulted path classification, the method shows an accuracy of 75,15\%. In contrast to the two previous location types, this result remains quite unsatisfying for fault location purposes.

\subsection{Relevancy of Multiple Artificial Neural Networks}

To justify and show the interest of dividing the prediction problem into smaller groups, and hence, using several \acrshort{ann}s, another method, that only differs at fault distance prediction and faulted path prediction stage where it uses a single \acrshort{ann} for each of the prediction, is tested. The results are compared with the multiple \acrshort{ann} developed in this paper in Table \ref{m1vsm2}.

\begin{table}
\renewcommand{\arraystretch}{1}
\footnotesize{
\begin{center}
\begin{tabular}{|c|c|c|}
\hline
{Prediction Type} & {Multiple ANN} & {Single ANN}\\
\hline
{Faulted Phase} & \multicolumn{2}{|c|}{100\%}\\
{Total Classification} & \multicolumn{2}{|c|}{}\\
\hline
{Fault Distance} & {0,40\%} & {2,49\%}\\
{Total Mean $e_{rel}$} & {} & {}\\
\hline
{Faulted Path} & {75,15\%} & {62,12\%}\\
{Total Classification} & {} & {}\\
\hline
\end{tabular}
\caption{ANNs prediction results}
\end{center}
}
\label{m1vsm2}
\end{table}

According to Table~\ref{m1vsm2}, one can easily conclude that the use of several \acrshort{ann}s is very beneficial, the multiple ANN method outstanding by far the single ANN method. Indeed, the fault distance prediction average relative error increases consequently to 2,49\% whereas the faulted path prediction accuracy is decreased down to 62,12\%. 
Concerning faulted phase classification, since both methods use a single and hence same \acrshort{ann}, the 100\% accuracy holds for both.

\section{Discussion}\label{sec:discuss}
As a result, one can conclude that the use of a set of \acrshort{ann} increases the overall accuracy of the method and lead to a competitive fault distance location. However, the major drawbacks of the method should be mentioned. 

\subsection{Robustness of the Method}\label{robust}

Since faults can occur under a quasi-unlimited number of scenarios as well as load values, the ability of the method to show good performance for unknown datasets, that is, its robustness, is a major concern. To study the robustness of the method, the method is tested on two new datasets: 
\begin{itemize}
    \item Dataset 1: of 532 samples with new fault scenario (\%DG: 30\%; $Z_f$: 0,5 and 5$\Omega$; $\theta_i$: 70° and 110°).
    \item Dataset 2: of 532 samples with similar fault scenario (\%DG: 10\%; $Z_f$: 0,1 and 1$\Omega$; $\theta_i$: 45° and 90°) but with load values increased of 30 \%.
\end{itemize}
In this paper, the choice has been made not to consider load variation for the method implementation, but testing the effect of load variation on the method would give an interesting insight into load generalization capacity. 
As a result, the method performance is significantly decreased in both cases as shown in Table~\ref{resultfl}, especially in terms of fault distance prediction going down to 3,95\% and 12,24\% for dataset 2 and 1 respectively. The method appears to be less affected by the load variation than by new fault scenarios. Nevertheless, path classification isn't as much impacted with 57,89\% and 63,53\% accuracy for new fault scenario and changed system load dataset respectively. Concerning faulted phase determination, the method shows great robustness, with results still very close to 100\%.

\begin{table}
\footnotesize{
\begin{center}
\begin{tabular}{|c|c|c|c|}
\hline
\multirow{2}{2em}{Results} & {Original} & {Dataset} & {Dataset}\\
 & {Dataset} & {1} & {2}\\
\hline
{Number of Faults} & {6384} & {532} & {532}\\
\hline
{Faulted Phase} & {100\%} & {98,87\%} & {100\%}\\
\hline
{Fault Distance} & {0,40\%} & {12,24\%} & {3,95\%}\\
\hline
{Faulted Path} & {75,15\%} & {57,89\%} & {63,53\%}\\
\hline
\end{tabular}
\caption{Results of fault location method application on different datasets}
\end{center}
}
\label{resultfl}
\end{table}

\subsection{Prediction Error Correlation}

The multiple \acrshort{ann}s strategies leads to potential correlation errors between the different prediction steps. A too-large error in fault distance prediction would lead to the use of the wrong \acrshort{ann}s for faulted path classification. Similarly, a wrong faulted phase prediction would lead to the use of the wrong input features for the rest of the workflow. For the sake of clarity, these two phenomena are called \acrfull{pce} and \acrfull{phce}, respectively. To observe the impact of these errors, the correlation errors impact on datasets 1 and 2 are shown in Table~\ref{errcorimpact}.

\begin{table}
\footnotesize{
\begin{center}
\begin{tabular}{|m{3.3cm}|c|c|}
\hline
{Results} & {Dataset 1} & {Dataset 2}\\
\hline
{Dataset samples} & {532} & {532}\\ [3.5ex]
\hline
\hline

{\% of \acrshort{phce}} & {1,13\%} & {0\%}\\
\hline
{\% of \acrshort{pce}} & {12,60\%} & {8,46\%}\\
\hline
{\% of \acrshort{phce} leading to \acrshort{pce} \hspace{1.5cm} / Number of cases} & {83,33\% / 6} & {n/a}\\ [3.5ex]
\hline
\hline

{Mean $e_{rel}$ in case of \acrshort{phce} \hspace{1.5cm} / Number of cases} & {31,02\% / 6} & {n/a}\\
\hline
{Potential mean $e_{rel}$ without \acrshort{phce} \hspace{.5cm} / Number of cases} & {12,02\% / 526} & {n/a}\\ [3.5ex]
\hline
\hline

{Faulted path accuracy in case of \acrshort{phce} / Number of cases} & {66,67\% / 6} & {n/a}\\
\hline
{Faulted path accuracy in case of \acrshort{pce} \hspace{.5cm} / Number of cases} & {25\% / 67} & {44,44\% / 45}\\
\hline
{Faulted path accuracy in case of both \acrshort{phce} and \acrshort{pce} / Number of cases} & {60\% / 5} & {n/a}\\
\hline

{Potential faulted path accuracy without correlation errors / Number of cases} & {62,5\% / 464} & {65,30\% / 487}\\
\hline
\end{tabular}
\caption{Impact of correlation errors on fault location for dataset 1 and 2}
\end{center}
}
\label{errcorimpact}
\end{table}

One can first observe that the direct impact on the distance prediction accuracy is quite significant, with a mean $e_{rel}$ up to 31,02\% in the case of dataset 1. In the case of faulted path classification, the accuracy goes down to 25\% and 44,44\% in datasets 1 and 2, respectively. Indeed, the overall impact of correlation errors is largely reduced by the small occurrence rate of these errors even for datasets 1 and 2. \acrshort{phce} occurs for only 1,13\% of the cases for dataset 1, whereas \acrshort{pce} occurrence is higher with 12,60\% and 8,46\% for dataset 1 and 2 respectively and directly related to $e_{rel}$. As a result, the faulted path classification is decreased from 62,5\% down to 57,89\% and from 65,30\% down to 63,53\% for dataset 1 and 2, respectively, whereas the relative distance error is slightly increased from 12,02\% to 12,24\% for dataset 1. It is also worth mentioning that most \acrshort{phce}s lead to \acrshort{pce}s as well, about 83\% in the case of dataset 1.

Due to these observations, one can conclude that even if the method is not affected by correlation errors, a better strategy on how to divide the faulted path prediction problem could be reviewed for potentially better results.

\subsection{Limitations}

As the proposed method relies on simulation, it follows well-known systems simulation best practices. Indeed, any time the real system changes, the simulation model should be updated accordingly, the dataset generated once again, and the \acrshort{ml} model retrained.

Additionally, the root of this method relies on the distribution system numerical model fidelity, in other words, how much the simulated faulted three-phase voltages mirror the real faults situation. Unpredictable elements such as interference, device measurement errors, and other disturbances cannot be simulated and considered by the software, which can consequently impact the method's accuracy.

Finally, the dataset relevancy stands as a major drawback. Faults labeled with path 1, due to its greater length, represent more than 53\% of the dataset, whereas path 3, which is very short, only represents 5\% of the dataset. This unbalanced training dataset leads the network to overfit on the most represented category, that is, path 1, and hence inaccurate for prediction purposes. Moreover, the use of specialized \acrshort{ann}s significantly reduces the size of the dataset that they can use. Indeed, the faulted path classification \acrshort{ann}s can only use as few as 408 or 504 of the 6384 samples generated, as shown in Table~\ref{accANN}. It is an important limitation knowing that the size of the dataset is an important criterion for building robust and accurate \acrshort{ann}s. However, the two former drawbacks aren't directly associated with the method itself since new datasets can be generated on demand.

\subsection{Method Relevancy}

The motivation for studying a new approach to fault location is to overcome or reduce the limitations of classical methods as well as improve their accuracy.

At first sight, this method would need measurement devices such as transducers only at the system substation. Unlike \acrshort{twbfl} methods, high-frequency measurement isn't necessarily required since the wavelet coefficients used here cover a frequency band in the range of kHz, easing costs for measurement devices. 

Accurate knowledge about the network and associated components is also required, similarly to \acrshort{ibfl} methods. This method stands out in its methodology, which remains the same for any kind of system and topology. Moreover, the short execution time of the method, once the machine learning model is trained, is appreciable, with the algorithm running in less than half a second given the three-phase voltages raw data. 

Finally, although at distribution levels \acrshort{ibfl} methods are the most accurate nowadays, the proposed method shows competitiveness in terms of accuracy, not to mention the potential along with future improvements.  

\subsection{Towards Improvements}

The main weakness of the method is its lack of robustness. A larger and more diversified dataset appears as a natural solution to reduce this weakness. In addition, the original dataset remains small when it comes to specialized \acrshort{ann}s that only use a part of this dataset. Finally, unbalanced datasets also negatively influence the overall accuracy of faulted path classification. To this extent, the first step toward the improvement of the method accuracy would be to retrain the related \acrshort{ann}s with a more complete, balanced, and larger dataset as well as to improve the faulted path location \acrshort{ann} set structure 
on of a more balanced (or a different) dataset.

Besides data quality, other factors can affect the accuracy of a given \acrshort{ann}, such as their hyper-parameters or their number of layers. An \acrshort{ann} structure remains basic compared to the power of deep learning. In that sense, trials could be done on more complex \acrshort{ann}. 

Finally, the accuracy of the distribution system modeling plays a key role in the method, which is why further work should include frequency-dependent component modeling, such as load and power lines, to gain model veracity and, at the same time, observe the impact on simulation time, an important factor for large dataset generation.

\section{Conclusion}\label{sec:conclude}
In this paper, a data-driven ground-fault location method in a distribution power system is explained and tested. The transient faulted three-phase voltage waveforms at the substation contain implicit information on fault location. These signals are processed using discrete wavelet transform. Mathematical statistic features of the wavelet coefficients, energy content, and three-phase faulted voltages are then calculated. 

We show that these features can help predict the fault location with a relative error of 0,4\% by using a workflow of several \acrshort{ann}s. Faulted phases are classified with an accuracy of 100\% whereas results for faulted path classification are lower with 75,15\% correct prediction. 

Even though the obtained results are promising and show the potential of the method, there is a need for future work to improve the method's accuracy and robustness. In particular, we plan to generate a larger and more diversified dataset, as well as explore other faulted path classification \acrshort{ann} workflow.
 
Finally, leveraging the results obtained in this paper, future research directions will focus on the development of digital twin solutions able to deal with real-life issues, such as the potential interferences and disturbances that can occur in the distribution system, and limit the method implementation.

\section{Acknowledgments}
This research is partially funded by the Swedish Knowledge Foundation, Grant No. 20150088 ``Software Technology for Self-Adaptive Systems'' and Grant No. 20200117 ``Aligning Architectures for Digital twin of the Organization''.


\appendix
\section{Distribution Power System Data}\label{ap:DPSdata}

\begin{table}[h]
\footnotesize{
\begin{center}
    \caption{Generation data of test distribution power system}
    \begin{tabular}{|c|}
\hline
Infinite Grid\\
\hline
$V_G=20$ kV, $f=60$ Hz, $X/R$ ratio $=10$\\
\hline
Synchronous Generator\\
\hline
$V_{DG}=3,5$ kV, $f=60$ Hz, $P_{nominal}=20$ MVA\\
\hline
    \end{tabular}
    \label{DataSource}
\end{center}
    }
\end{table}


\begin{table}[h]
\footnotesize{
\begin{center}
\caption{Transformers data of test distribution power system}
\begin{tabular}{|c|c|c|c|}
\hline
\multirow{3}{6em}{Transformers} & {Nominal} & \multirow{3}{4em}{Voltage} & \multirow{3}{5em}{Connection} \\
& {Power} &  &\\ 
& {(MVA)} & {(kV)} &\\
\hline
{Infinite Grid-Bus 1} & {60} & {63/20} & {Yg-Yg}\\
\hline
{Bus 8-DG} & {10} & {3.5/20} & {Yg-Yg}\\
\hline
{Bus 3-Load 3} & {10} & {20/0,4} & {Yg-Yg}\\
\hline
{Bus 4-Load 4} & {10} & {20/0,4} & {Yg-Yg}\\
\hline
{Bus 6-Load 6} & {10} & {20/0,4} & {Yg-Yg}\\
\hline
{Bus 8-Load 8} & {10} & {20/0,4} & {Yg-Yg}\\
\hline
{Bus 10-Load 10} & {10} & {20/0,4} & {Yg-Yg}\\
\hline
{Bus 11-Load 11} & {10} & {20/0,4} & {Yg-Yg}\\
\hline
\end{tabular}
\end{center}
}
\label{DataTransfo}
\end{table}


\begin{table}[h]
\footnotesize{
\begin{center}
\caption{Loads data of test distribution power system}
\begin{tabular}{|c|c|c|c|c|}
\hline
{Load} & {Model} & {Connection} & {P} & {Q} \\
{Bus} & & & {(MW)} & {(kVar)}\\
\hline
{3} & {Static, constant Z} & {Yg} & {7,5} & {42,5}\\
\hline
{4} & {Static, constant Z} & {Yg} & {7,5} & {42,5}\\
\hline
{6} & {Static, constant Z} & {Yg} & {3,6} & {12,5}\\
\hline
{8} & {Static, constant Z} & {Yg} & {7,5} & {42,5}\\
\hline
{10} & {Static, constant Z} & {Yg} & {7,5} & {42,5}\\
\hline
{11} & {Static, constant Z} & {Yg} & {3,6} & {12,5}\\
\hline
\multicolumn{3}{|c|}{Total} & {37.2} & {195}\\
\hline
\end{tabular}
\end{center}
}
\label{DataLoads}
\end{table}

\begin{table}[h]
\footnotesize{
\begin{center}
\caption{Power lines data of test distribution power system}
\begin{tabular}{|c|c|c|c|}
\hline
{Line} & {DC} & {Inner} & {Outer}\\
{Type} & {Resistance $(\Gamma/km)$} & {Radius} & {Radius}\\
 & {$(\Gamma/km)$} & {(cm)} & {(cm)}\\
\hline
{J Marti} & {0,065} & {0,14} & {0,8}\\
\hline
{Path} & \multicolumn{2}{|c|}{Line between Buses} & {Length (km)}\\
\hline
{1} & \multicolumn{2}{|c|}{1-2} & {2}\\
\hline
{1} & \multicolumn{2}{|c|}{2-5} & {1,5}\\
\hline
{1} & \multicolumn{2}{|c|}{5-7} & {1,5}\\
\hline
{1} & \multicolumn{2}{|c|}{7-9} & {3}\\
\hline
{1} & \multicolumn{2}{|c|}{9-11} & {2}\\
\hline
{2} & \multicolumn{2}{|c|}{2-3} & {2}\\
\hline
{3} & \multicolumn{2}{|c|}{2-4} & {1}\\
\hline
{4} & \multicolumn{2}{|c|}{5-6} & {1}\\
\hline
{5} & \multicolumn{2}{|c|}{7-8} & {2}\\
\hline
{6} & \multicolumn{2}{|c|}{9-10} & {3}\\
\hline
\end{tabular}
\end{center}
}
\label{DataLine}
\end{table}


\begin{figure}[h]
\begin{center}
\includegraphics[scale=0.5]{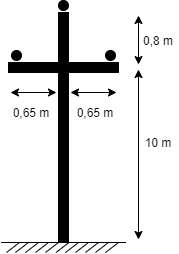}
\end{center}
\caption{Power lines model of test distribution power system.}
\label{Line format}
\end{figure}




\begin{table*}[h]
\footnotesize{
\begin{center}
\caption{Relevant features for optimized workflow}
\begin{tabular}{|m{3cm}|m{8.3cm}|}
\hline
{ANNs} & {Relevant Features}\\ 
\hline 
{Tfp} & {$std(V_a)$, $std(V_b)$, $std(V_c)$, $std(V_aCD_8)$, $std(V_bCD_8)$, $std(V_cCD_8)$, $std(V_aCA_8)$, $std(V_bCA_8)$, $std(V_cCA_8)$, $E_{wave}(V_aCD_8)$, $E_{wave}(V_bCD_8)$, $E_{wave}(V_cCD_8)$, $E_{wave}(V_aCA_8)$, $E_{wave}(V_bCA_8)$, $E_{wave}(V_cCA_8)$}\\ [2ex]
\hline 
{Ofd-i} & {$var(V_i)$, $var(V_iCD_8)$, $skn(V_i)$, $skn(V_iCD_8)$, $skn(V_iCA_8)$, $cm3(V_i)$, $cm3(V_iCD_8)$, $cm3(V_iCA_8)$, $E_{wave}(V_iCD_8)$}\\ [2ex]
\hline 
{Ofd-ij} & {$var(V_iCD_8)$, $var(V_jCD_8)$, $skn(V_i)$, $skn(V_j)$, $skn(V_iCA_8)$, $skn(V_jCA_8$, $cm3(V_i)$, $cm3(V_j)$, $cm3(V_iCD_8)$, $cm3(V_jCD_8)$, $cm3(V_iCA_8)$, $cm3(V_jCA_8)$, $mode(V_iCD_8)$, $mode(V_jCD_8)$, $E_{wave}(V_iCD_8)$, $E_{wave}(V_jCD_8)$}\\ [2ex]
\hline 
{Ofd-ABC} & {$var(V_ACD_8)$, $var(V_BCD_8)$, $var(V_CCD_8)$, $var(V_1CD_8)$, $var(V_2CD_8)$, $var(V_1CA_8)$, $var(V_2CA_8)$, $skn(V_A)$, $skn(V_B)$, $skn(V_C)$, $skn(V_1)$, $skn(V_ACA_8)$, $skn(V_BCA_8)$, $skn(V_CCA_8)$, $skn(V_1CA_8)$, $skn(V_2CA_8)$, $cm3(V_A)$, $cm3(V_B)$, $cm3(V_C)$, $cm3(V_1)$, $cm3(V_ACD_8)$, $cm3(V_BCD_8)$, $cm3(V_CCD_8)$, $cm3(V_ACA_8)$, $cm3(V_BCA_8)$, $cm3(V_CCA_8)$, $cm3(V_1CD_8)$, $cm3(V_2CD_8)$, $mode(V_ACD_8)$, $mode(V_BCD_8)$, $mode(V_CCD_8)$, $mode(V_ACA_8)$, $mode(V_BCA_8)$, $mode(V_CCA_8)$, $mode(V_2CA_8)$, $E_{wave}(V_ACD_8)$, $E_{wave}(V_BCD_8)$, $E_{wave}(V_CCD_8)$, $E_{wave}(V_ACA_8)$, $E_{wave}(V_BCA_8)$, $E_{wave}(V_CCA_8)$, $E_{wave}(V_1CD_8)$, $E_{wave}(V_2CD_8)$, $E_{wave}(V_1CA_8)$, $E_{wave}(V_2CA_8)$}\\ [2ex]
\hline 
{Ofp-i$H_1$ \& Ofp-i$H_2$} & {$var(V_iCD_8)$, $var(V_iCA_8)$, $skn(V_i)$, $skn(V_iCD_8)$, $cm3(V_i)$, $cm3(V_iCD_8)$, $mode(V_i)$, $mode(V_iCA_8)$}\\ [2ex]
\hline 
{Ofp-ij$H_1$ \& Ofp-ij$H_2$} & {$var(V_i)$, $var(V_j)$, $var(V_iCD_8)$, $var(V_jCD_8)$, $var(V_iCA_8)$, $var(V_jCA_8)$, $skn(V_i)$, $skn(V_j)$, $skn(V_iCD_8)$, $skn(V_jCD_8)$, $cm3(V_i)$, $cm3(V_j)$, $cm3(V_iCD_8)$, $cm3(V_jCD_8)$, $cm3(V_iCA_8)$, $cm3(V_jCA_8)$, $mode(V_iCD_8)$, $mode(V_jCD_8)$, $E_{wave}(V_iCD_8)$, $E_{wave}(V_jCD_8)$, $E_{wave}(V_iCA_8)$, $E_{wave}(V_jCA_8)$}\\ [2ex]
\hline 
{Ofp-ABC$H_1$ \& Ofp-ABC$H_2$} & {$var(V_2CD_8)$, $skn(V_A)$, $skn(V_B)$, $skn(V_C)$, $skn(V_0)$, $skn(V_1)$, $skn(V_ACD_8)$, $skn(V_BCD_8)$, $skn(V_CCD_8)$, $skn(V_1CD_8)$, $skn(V_2CD_8)$, $cm3(V_A)$, $cm3(V_B)$, $cm3(V_C)$, $cm3(V_0)$, $cm3(V_1)$, $cm3(V_ACD_8)$, $cm3(V_BCD_8)$, $cm3(V_CCD_8)$, $cm3(V_1CD_8)$, $cm3(V_2CD_8)$, $mode(V_ACD_8)$, $mode(V_BCD_8)$, $mode(V_CCD_8)$, $mode(V_2CD_8)$, $E_{wave}(V_2CD_8)$}\\
\hline
\end{tabular}
\end{center}
}
\label{featwf2}
\end{table*}

\end{document}